\documentclass[final,1p,times,twocolumn]{elsarticle}

 \usepackage{graphics}
 \usepackage{graphicx}
 \usepackage{epsfig}

\usepackage{amssymb}
\usepackage{amsthm}

\journal{New Astronomy}

\begin{document}

\begin{frontmatter}



\title{Probing clustering features around Cl 0024+17}


\author{A.L.B. Ribeiro}

\address{Center for Particle Astrophysics, Fermi National Accelerator Laboratory\\
Batavia, IL, 60510-0500, USA}
\address{Laborat\'orio de Astrof\'{\i}sica Te\'orica e Observacional\\ 
              Departamento de Ci\^encias Exatas
e Tecnol\'ogicas\\ Universidade Estadual de Santa Cruz -- 45650-000, Ilh\'eus-BA, Brazil\\}

\begin{abstract}
I present a spatial analysis of the galaxy distribution around the cluster Cl 0024+17.
The basic aim is to find the scales where galaxies present a significant deviation
from an inhomogeneous Poisson statistical process.
Using the generalization of the Ripley, Besag, and
the pair correlation functions for non-stationary point patterns, I estimate these transition scales
for a set of 1,000 Monte Carlo realizations of the Cl 0024+17 field, 
corrected for completeness up to the outskirts.
The results point out the presence of at least two physical scales in this field
at $31.4^{\prime\prime}$ and $112.9^{\prime\prime}$. The second one is statistically
consistent with the dark matter ring radius ($\sim 75^{\prime\prime}$) previously identified by
Jee et al. (2007). However, morphology
and anisotropy  tests point out that a clump at $\sim 120^{\prime\prime}$ NW
from the cluster center could be the responsible for the second transition scale.
These results do not indicate the existence of a galaxy counterpart
of the dark matter ring, but the methodology developed to study the galaxy field as
a spatial point pattern provides a good statistical evaluation of the physical scales around the cluster. I briefly discuss the usefulness of this approach
to probe features in galaxy distribution and N-body dark matter simulation data.
\end{abstract}


\end{frontmatter}


\section{Introduction} 

Galaxy clusters are the largest virialized structures in the universe.
These systems are composed of three components behaving differently
during collisions: galaxies, hot gas and dark matter.
While the hot intergalactic gas presents strong eletromagnetic interactions
during  clusters enconteurs, both dark matter and galaxies are predicted to
be collisioness, see e.g. \cite{marke1}. Hence, clusters are important
laboratories for studying the dynamics of the different kinds of matter
in the universe. 
Indeed, galaxy clusters are themselves the result of several
mergers of smaller groups of galaxies, see e.g. \cite{davis85}, a process still ongoing
in many systems like, for instance, the so-called 'Bullet Cluster' (1E 0657-56) 
[\cite{mark2,clo1}], and other clusters like Cl 0152-1357 and MS 1054, see \cite{jee05a,jee05b}.
Also, cluster Cl 0024+17 has been a target of many studies since its discovery
by \cite{humsan}. This is an intermediate-redshift system (z=0.395) with both
weak and strong lensing \cite{tys98,broad00,comer06}. There are several 
indications that this cluster, an apparently relaxed system, is actually the
portrait of a collision of two clusters, along our line of sight 
\cite{czo1,czo2,ota,zhang,jee07}. The weak lensing analysis presented by \cite{jee07}
shows a ring-like structure in the projected matter distribution at $r\sim 75^{\prime\prime}$
($\sim$ 0.4 Mpc), surrounding a soft, dense core at $r\leq 50^{\prime\prime}$.
The authors interpret this substructure as the result of a high-speed line-of-sight collision
of two massive clusters $\sim$1-2 Gyr ago. An interesting question in this context refers 
to the behaviour of the galaxy component with respect to dark matter.
In the case of the Bullet Cluster, Cl 0152-1357 and MS 1054-0321, the distribution
of galaxies follows dark matter quite well \cite{mark2,clo1,jee05a,jee05b}.
However, the recent work of \cite{qin}, based on the 2D distribution of galaxies,
suggests that the ringlike structure observed in 
dark matter measurements is not seen in the projected two-dimensional galaxy
distribution.

In the present work, I introduce an alternative approach to probe both in radial and angular
coordinates the projected
galaxy distribution around galaxy clusters, and applied it to the case
of Cl 0024+17. The basic aim is to find the scales where galaxies present a significant deviation from an inhomogeneous Poisson statistical process and compare that 
to the dark matter ring radius and the secondary clump present in the field.

\section{Galaxy Sample}

The sample of galaxies around Cl 0024+17 used in this work is taken from
the catalog of Czoske et al. (2001). This corresponds to 295 galaxies in the range
$0.37<z<0.41$, the most probable cluster members in the 
$\sim 0.3^\circ \times 0.3^\circ$ field. The center of the cluster was
chosen to be the geometric center of the dark matter ring, see \cite{jee07}.
Distance $r$ for each galaxy was calculated from their equatorial coordinates,
in arcsecs. Since the completeness of the sample varies from $>$80\% at the cluster
center to $<$50\% in the outer regions, the catalog was corrected
through 1,000 Monte Carlo realizations of the sample. This was done
following approximately the completeness variation map of
\cite{czo1}, adopting a procedure similar to \cite{qin}.
One example of such realizations is presented in Figure 1, where 
black points are taken from the simulation and green points
are the real galaxies from \cite{czo1}. The background in
this figure is a density map based on the void distance from each point
with respect to a fine grid \cite{bade,bade1}. This is just introduced to illustrate how the
simulations fill the empty cells left by the original sample.

\begin{figure}
\includegraphics[width=9.5cm,height=8.5cm]{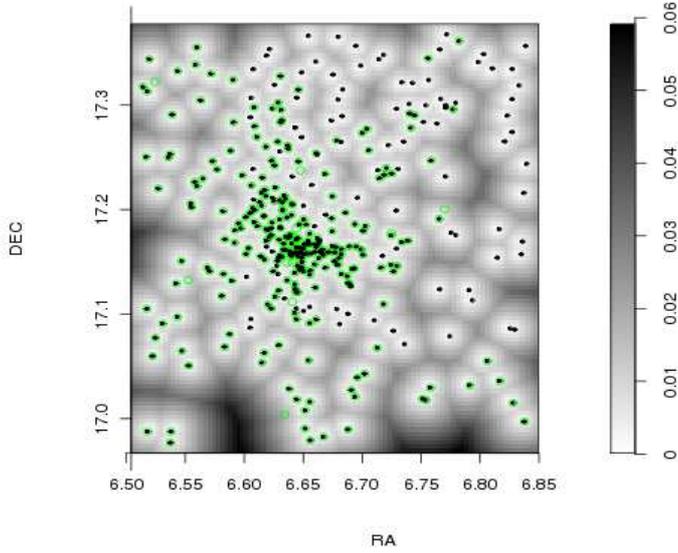}
\caption{Simulated + real field on a void distance map. Simulated
and real points are in black and green, respectively. The
background is a density image built from the void distance
matrix from each point with respect to a fine grid.\label{fig1}}
\end{figure}

\section{Methodology}

\subsection{Spatial Point Patterns}

Galaxy distribution around Cl  0024+17 is the point pattern we want to probe.
In statistical theory, a spatial point pattern (SPP) ${\bf x}=\{x_1,...,x_n\}$ is simply 
a collection of points $x_i$ in some
bounded area, and interpreted as a realization of a more generic point process ${\bf X}$.
Analysis and modelling of SPP is a challenging topic in modern spatial spatistics.
The mathematical theory was first developed to solve various
problems where it is sensible to model the locations of events as random, \cite{illian}.
Recently, statistical theory
has made many developments with respect to SPP analysis. One of these advances is 
related to achievements in the study of nonstationary processes. Statistical 
stationarity (or homogeneity) means
that the distribution of the point process is invariant under translations.
This implies that the intensity $\lambda$, i.e. the expected number of points per unit
area of the SPP, is constant. Nonstationarity, otherwise, allows for deterministic
local variation of the intensity:$\lambda\rightarrow\lambda (u)$, where
$u$ is a given location on the SPP. This is the case we are interested in this
work, since galaxy distribution in the sampling window around the cluster
is obviously inhomogeneous.

\subsection{Ripley K-function}

In spatial statistics, Ripley K-function \cite{rip} is a classical tool to analyse SPP.
It is defined as the expected number of events 
within the distance $r$ of any given point divided by the intensity $\lambda$. The K-function, being
a measure of the distribution of the inter-point distances,captures the spatial dependence between
different regions of a SPP. Mathematically, the inhomogeneous K-function is defined as

\begin{equation}
K_{inhom}(r)={\rm E}\left[{1\over \lambda (u)}\sum_{x_j\in {\bf\rm X}}{1\over\lambda (x_j)}
{\bf 1}\{0<||u-x_j||\leq r\}~|~ u\in {\bf\rm X}\right].
\end{equation}

\noindent Thus, $\lambda(u)K(r)$ is the expected total 'weight' of all random points within
a distance $r$ of the point $u$, where the weight of a point $x_i$ is $1/\lambda(x_i)$ 
\cite{bade2}, and for a Poisson process with intensity function
$\lambda(u)$, the K-function reduces to $K_{inhom}=\pi r^2$.

A standard computational estimator of K is given by

\begin{equation}
\hat{K}_{inhom}(r)={1\over area(W)}\sum_i\sum_{j\ne i}{{\bf 1}\{||x_i-x_j||\leq r\}\over
\hat{\lambda}(x_i)\hat{\lambda}(x_j)}e(x_i,x_j;r),
\end{equation}

\noindent where $e(x_i,x_j;r)$ is an edge correction weight, $W$ is the
observational window, and $\hat{\lambda}(u)$ is an 
estimate of the intensity function $\lambda(u)$.

Although the K-function is the classical tool to probe SPP, modern spatial statistics
rarely uses K(r), but rather its variant, the L-function as introduced by \cite{besag}.
In the planar case, the L-function estimator is defined as

\begin{equation}
\hat{L}_{inhom}(r)=\sqrt{{\hat{K}_{inhom}(r)\over \pi}}.
\end{equation}

\noindent The advantage of using $L(r)$ is that it is always proportional to $r$, and in 
the Poisson case $L(r)=r$,  \cite{illian}. Thus, $L(r)-r>0$ indicates
clustering with respect to a Poisson process.

In this work, the L-function was applied to the galaxy distribution
around Cl 0024+17 using the estimation algorithms
provided in the library {\bf spatstat}, developed by \cite{bade}, and running under the R statistical package
(www.cran.r-project.org). The edge correction used here 
is implemented in the tasks {\bf Kinhom} and {\bf Linhom}. It is
briefly described in the next.

\subsection{Edge Corrections}

Edge corrections are used to correct biases in the estimation of K(or L)-function
due to the loss of information near to the boundary of $W$. 
In this work, I have applied  the following edge correction:

\begin{equation}
e(x_i,x_j;r)= {{\bf 1}(b_i > r)\over{\sum_j {\bf 1}(b_j > r){1\over\lambda(x_j)}}},
\end{equation}

\noindent where $b_i$ is the distance from $x_i$ to the boundary of the
window $W$. Among other options available in 
{\bf spatstat}, this weighting function is considered the most computationally and
statistically efficient when there are large numbers of points, see \cite{bade2}.

\subsection{Intensity estimator $\hat{\lambda}(u)$}

Finally, there remains the question of how to estimate the intensity function $\lambda(u)$.
This is estimated using a `leave-one-out' kernel smoother, as described in \cite{bade2}.  
The estimate $\hat{\lambda}(x_i)$ 
 is computed by removing $x_i$ from the point pattern, then
applying a Gaussian kernel smoothing to the remaining points, 
and finally evaluating the smoothed intensity at $x_i$ (in this work
we set the kernel width equal to $30^{\prime\prime}$).
This estimator is also included in {\bf Kinhom} and {\bf Linhom}.

\section{Analysis of Cl 0024+17}

I have used the routine {\bf Linhom} to probe the galaxy distribution
around Cl 0024+17. For each MC realization of this field, I found the
scales where clustering weakens towards an inhomogeneous
Poisson process. In Figure 1, regions above the line $L(r)=r$
describes clustered distribution of points. To test the null hypothesis of
an inhomogeneous Poisson process I also plot in this figure the 95\% envelopes
computed from $M$ independent simulations of this process, with the
same intensity as the galaxy field, i.e.,
$\hat{L}^{(j)}(r)$ for $j=1,...,M$. The upper and lower envelopes of these simulated
curves are

\begin{equation}
\hat{L}^{lower}(r)=\min_j\hat{L}^{(j)}(r)
\end{equation}

\begin{equation}
\hat{L}^{upper}(r)=\max_j\hat{L}^{(j)}(r).
\end{equation}

\noindent If data come from a Poisson process, then $\hat{L}(r)$ and
$\hat{L}^{(1)}(r)$,...,$\hat{L}^{(M)}(r)$ are statistically equivalent and independent, so
the probability that $\hat{L}(r)$ lies outside the envelopes is equal to
$2/(M+1)$ by symmetry. This corresponds to the significance level $\alpha$ in the test
which rejects the null hypothesis. Hence, to compute the 95\% envelopes ($\alpha=0.05$), we need only
39 MC simulations of the Poisson process, see e.g. \cite{bade1}.

In Figure 2, we see two different regions where the null hypothesis of a Poisson process
is rejected at the 5\% significance level. The transition scales
where the clustered process weakens towards a Poisson process are marked
with green dashed lines. A single transition around a cluster indicates its outer limit, the
region where galaxy distribution declines in clustering (or intensity)
and turns indistinguishble from a Poisson process.
Two (or more) transitions indicate a recoverage of the clustering process. Any 
morphological feature could
be responsible for that: clumps, rings, tails, etc. In the present analysis,
about 88\% of the MC simulations
present two scales like those we see in Figure 2. 
The Shapiro-Wilk's test fails to detect that the distributions of
${\rm 1^{st}}$ and ${\rm 2^{nd}}$ transition scales significantly deviate from normal
distributions: $\mu_1=31.4^{\prime\prime}; \sigma=8.4^{\prime\prime}$ (p-value=0.3603) and 
$\mu_1=112.9^{\prime\prime}; \sigma=21.4^{\prime\prime}$ (p-value=0.7338).
Also, Bartlett's and Kruskal-Wallis's tests strongly reject the hypotheses
of equal variances and means, respectively, for these two distributions (p-value $<2.2 \times 10^{-16}$ in both
tests). All these results indicate the presence of two distinct physical scales in
the Cl 0024+17 galaxy field, as we can see in Figure 3. Note that the
dark matter ring is located at $r\sim 75^{\prime\prime}$ \cite{jee07}, and separations between this
radius and the $1^{\rm st}$ and $2^{\rm nd}$ scales are equal to $\sim 5.17\sigma_1$
and $\sim 1.77\sigma_2$, respectively. This suggests that the dark matter ring
is more likely to be associated to the second region. Also in this figure, it is
presented the average correlation function for all set of MC simulations (see
the small box). Note that the $1^{\rm st}$ transition scale corresponds just to a narrow
diminishment interval in the correlation function, while the DM ring radius and the
$2^{\rm nd}$ transition scale are approximately coincidents with the highest peaks
of $\xi (r)$ after $\sim 50^{\prime\prime}$. Not by chance this scale roughly marks the
intersection of the two Gaussian distributions. These
results indicate that the system has a core at $\lesssim 50^{\prime\prime}$ (which agrees
with \cite{jee07}) and that there is an extended region after this radius
with two significant peaks ($\sim 3\sigma$ above $\bar{\xi}$).
One of them agrees well with the position of the DM ring radius while the other
coincides with the mean of the $2^{\rm nd}$ transition scale.

\begin{figure}
\includegraphics[width=8cm,height=8cm]{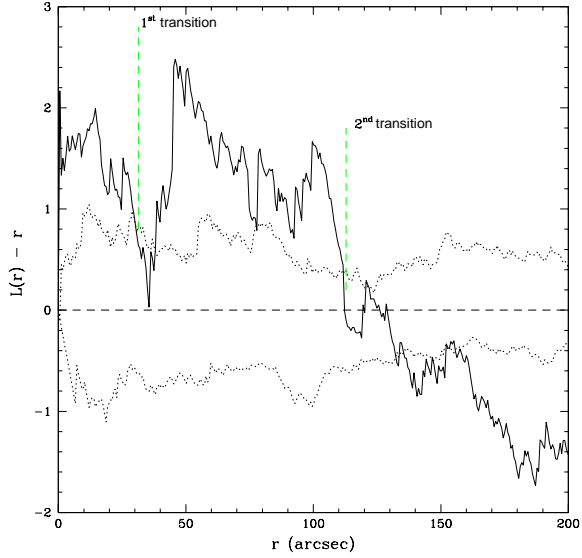}
\caption{Besag L function for one realization
of the Cl 0024+17 galaxy field (solid lines). The Poisson process corresponds to the
line L(r)=r (dashed lines). The 95\% lower and upper envelopes (averaged over
all the MC realizatios of the galaxy field) are presented in dotted lines.
Transition scales are marked with vertical green dashed lines\label{fig2}}
\end{figure}

\begin{figure}
\includegraphics[width=8cm,height=8cm]{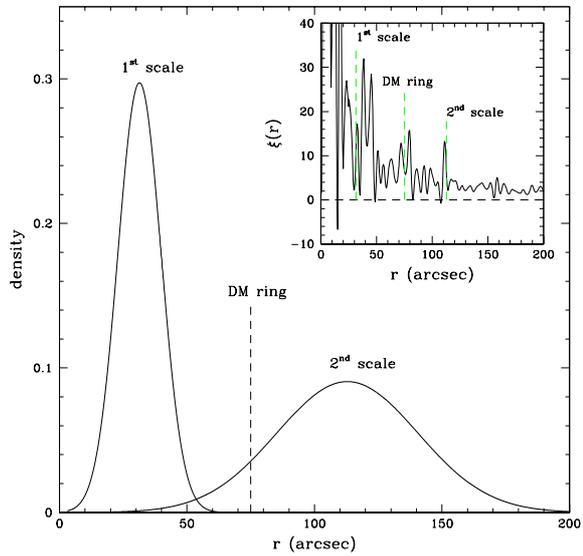}
\caption{Normal distributions of the $1^{\rm st}$ and $2^{\rm nd}$ transition scales.
Vertical dashed lines indicate the position of the DM ring radius. Small box: average correlation
function for all MC simulations. DM ring, $1^{\rm st}$ and $2^{\rm nd}$ scales
are indicated by green dashed lines.\label{fig3}}
\end{figure}

\section{Morphology test}

One should be careful not to overinterpret the previous analysis by
concluding that the second scale we see in Figure 3 implies that the galaxy distribution
around Cl 0024+17 has a luminous counterpart of the dark matter ring. 
Indeed, this result may have to do with the way observations were done.
For instance, the completeness map was constructed after a smoothing procedure using 
a Gaussian kernel of $30^{\prime\prime}$ \cite{czo2}. This size is similar
to that I have found for the first transition scale, but the separation between the
two normal distributions is $\sim$2.7 times this smoothing scale, so it seems that
the width of the kernel is not large enough to
invalidate the results. On the other hand, the presence of a
secondary peak in the distribution about $120^{\prime\prime}$ NW of the cluster center
(see Figure 1 and Figure 4 of \cite{czo2}) can be the responsible for the
second scale detected in data. This can be seen in Figure 4, where I plot the
perspective view of galaxy distribution also
smoothed by a Gaussian kernel of width $30^{\prime\prime}$. Letters A and B indicate the first and 
second density peaks in the field. Note that the peaks are very close to each other, although clearly distinct.
Less distinguishble, however, can be the effect of a spheroidal clump from a ring in the spatial analysis I have
applied to data. One way to measure such morphologic effects is now presented.

\begin{figure}
\includegraphics[width=8cm,height=10cm]{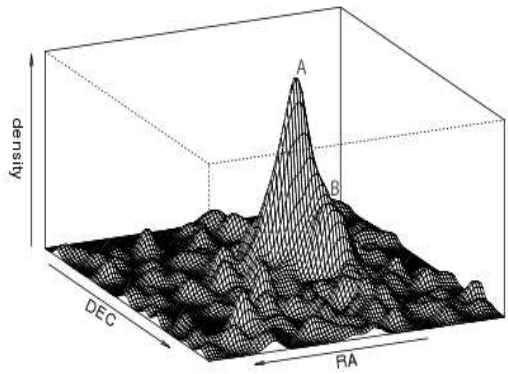}
\caption{Perspective view of galaxy distribution around Cl 0024+17
smoothed by a Gaussian kernel of width $30^{\prime\prime}$. Letters A and B indicate the the first and 
second density peaks in the field.
\label{fig4}}
\end{figure}

I have built two mock samples made to mimic the cases we want to study. In Model I,
we have a central cluster with radius equal to $\mu_1=31.4^{\prime\prime}$, plus
a clump at $\sim 120^{\prime\prime}$ NW (radius defined as $\mu_1/2$)
and a Poisson background with same intensity of
the Cl 0024+17 field. Both the cluster and the clump have a Hernquist profile (Hernquist 1990), but
the cluster is set to have $\sim 4.5$ times the density of the clump
(after the Gaussian smoothing). Model II is the central cluster plus a ring 
at the second scale, $112.9^{\prime\prime}$,with width equal to $\sigma_2=21.4^{\prime\prime}$. 
The density of the cluster is also $\sim 4.5$ times the density of the ring. For each case,
1,000 mock fields were generated. Examples are presented in Figure 5, pannels a and c.

Analysis of these mock fields shows that:
(i) in both cases two statistically distinct scales are found -- Bartlett and Kruskal-Wallis tests
reject the null hypotheses of equal variances and means at $>$99\% confidence (see resultant
normal distributions in Figure 5, pannels
b and d); (ii) however, while in Model II about 94\% of the realizations 
present two scales, this only happens for 61\% of the Model I realizations;
(iii) the $1^{\rm st}$ Gaussian is narrower for Model I 
($\mu_1^{\rm I}=30.1,\sigma_1^{\rm I}=7.7,\mu_1^{\rm II}=31.3,\sigma_1^{\rm II}=9.7$)  while the
$2^{\rm nd}$ Gaussian is narrower for Model II ($\mu_2^{\rm I}=117.3,\sigma_2^{\rm I}=29.0,\mu_2^{\rm II}=120.0,\sigma_2^{\rm II}=19.3$); 
(iv) however, there is no evidence for difference in the results of Models 1 and 2 --
Bartlett, ANOVA and Tukey tests indicate only a 0.001\% probability of difference between the samples.

The morphology test presented here is quite naive
in the sense that neither model reproduces the complexity of the real galaxy field. 
Exactly for this reason, it may indicate that the methodology presented in Section 3 is not accurate enough to discriminate different morphologies around the galaxy cluster. To improve the results, I have applied
a wavelet analysis to the galaxy field of Cl 0024+17.  Here I followed the method of 
\cite{sbm}.
Basically, the coordinates of the galaxies are convolved by a radial Mexican Hat filter
on a grid of $N\times N$ pixels. The method leads to a multiscale analysis
which starts below the biggest scales, where the whole cluster is detected as a unique structure,
and ends with the smallest scale, when approximately only one galaxy lies  into the operating area
of the wavelet. To obtain the significance level of a given structure I have applied the
replica procedure, where many simulations are made by drawing independently $X_i$ and $Y_i$
from the the $X$ and $Y$ distributions of the sample studied to destroy small-scale correlations, see e.g. \cite{mazu}.
The analysis is then performed on the "random fields" for the same wavelet scale than the true field.
The significance level of a given structure is estimated from the number of images $N_{ext}$ in which 
maxima greater than the maxima wavelet coefficients are found among the whole set of simulations $N_{set}$. 
Thus, $P_{SL}=N_{ext}/N_{set}$, see \cite{mazu}.
In Figure 6 we present the isosurfaces of the wavelet coefficients corresponding to the scales
$a=64,32,16,~{\rm and}, 8$, respectively (in pixels units for a map of $128\times 128$ pixels).
The wavelet scale used is $a=64$ pixels corresponding to 1.0 Mpc. In the top left low resolution image
($a=64$ pixels), we see the prominent central region of the cluster with some density enhament around it, which could suggest an almost symmetric substructure at first sight. 
However, with increasing resolution, this feature disappears, while
at scales $a=32~{\rm and}~64$ a core + clump structure emerges. Finally, at
scale $a=8$ (bottom right image) just a central peak remains. For the two intermediate
scales I estimate the significance level of a model core + clump, having found
$P_{SL}=0.023$ ($a=32$) and $P_{SL}=0.045$ ($a=16$) (for 1,000 MC simulations), 
which suggests that such a model cannot be rejected at these scales.

\begin{figure}
\includegraphics[width=6.0cm,height=6.0cm,angle=0]{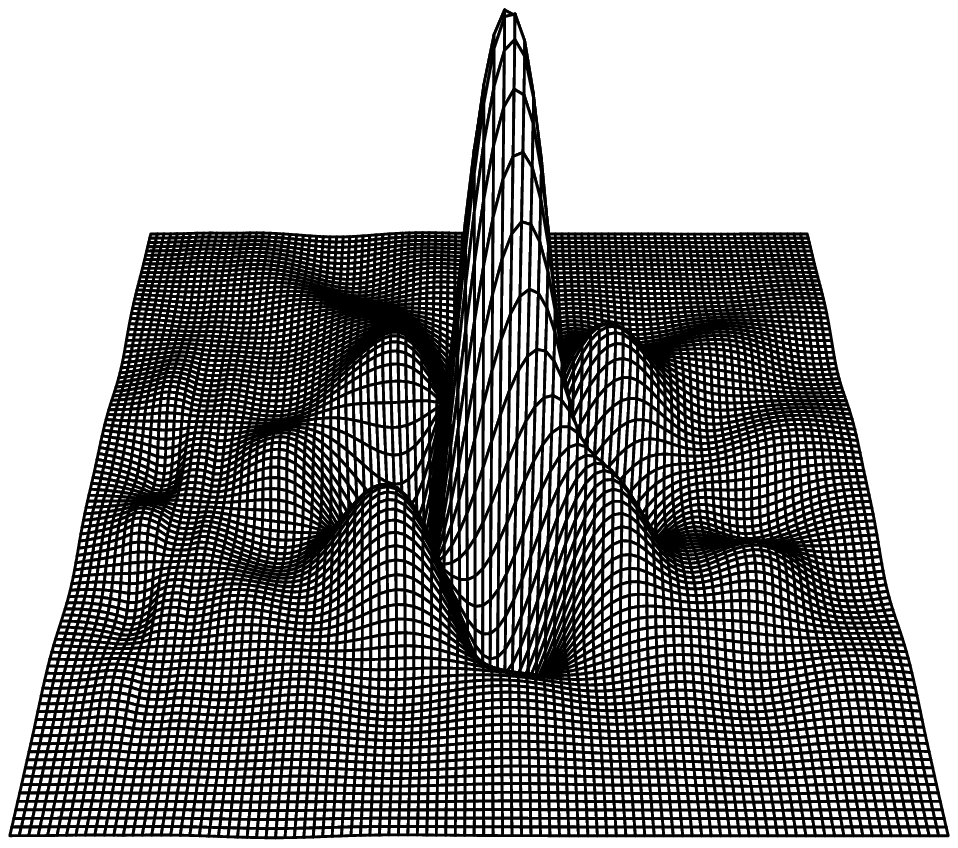}
\includegraphics[width=6.0cm,height=6.0cm,angle=0]{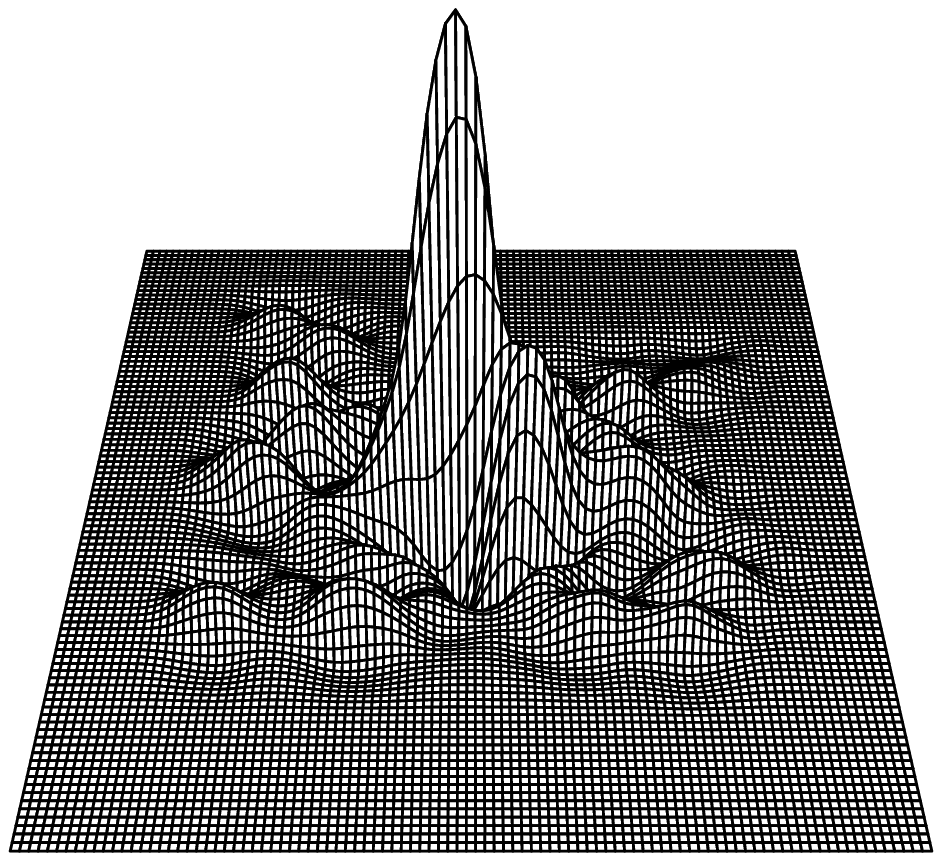}
\includegraphics[width=6.0cm,height=6.0cm,angle=0]{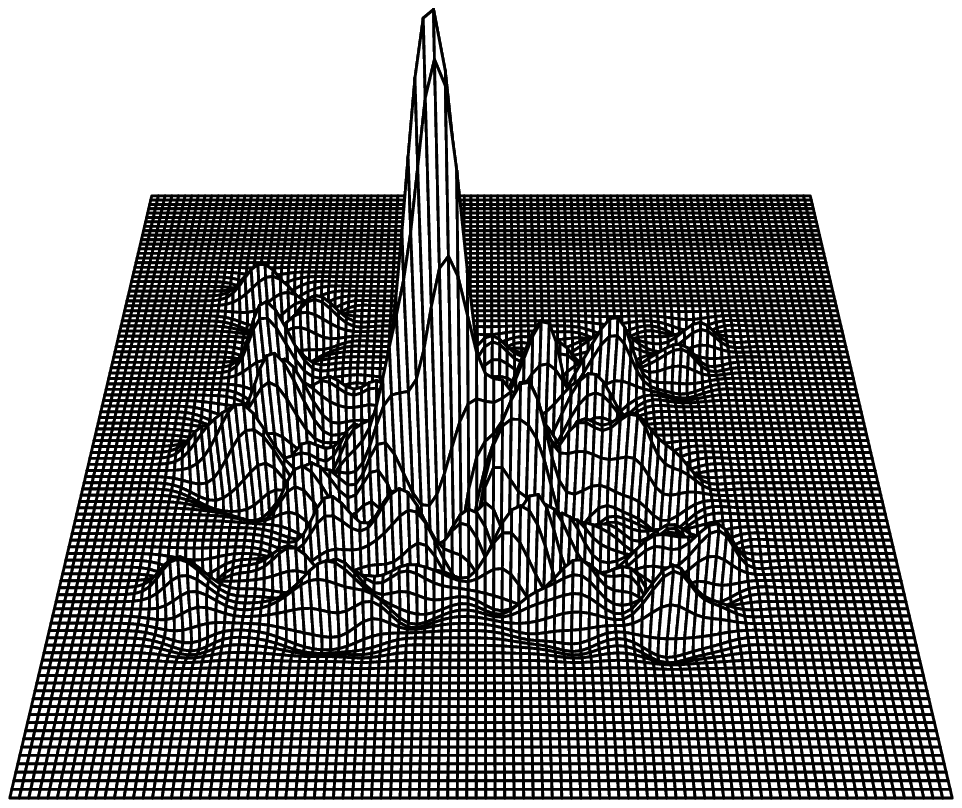}
\includegraphics[width=6.0cm,height=6.0cm,angle=0]{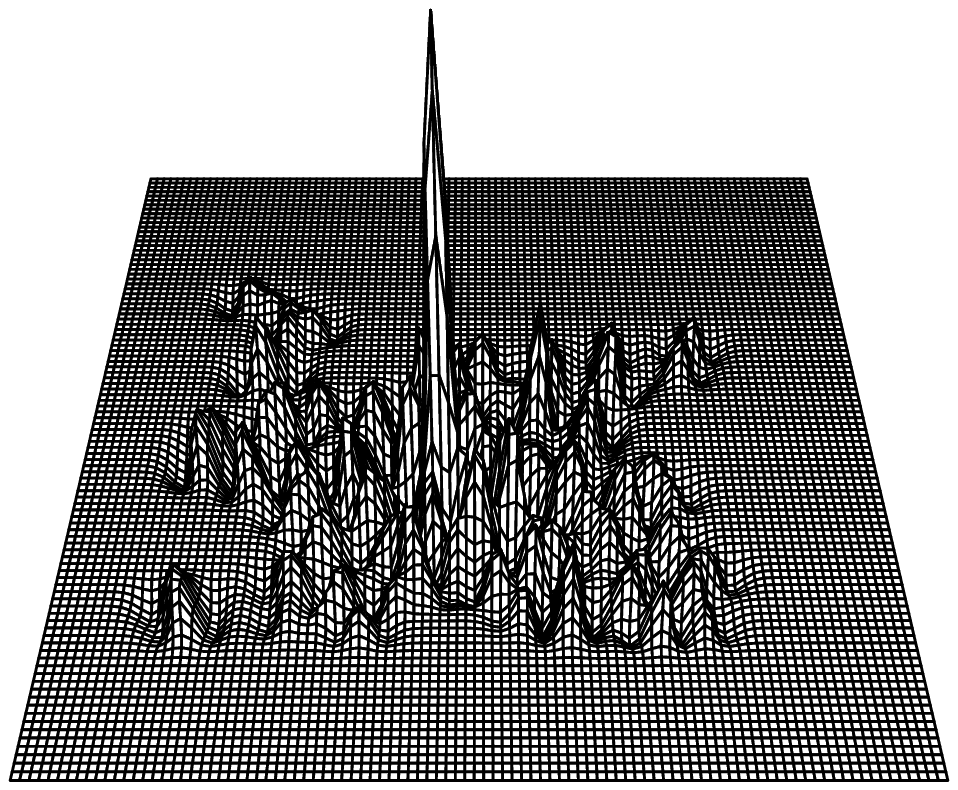}
\caption{Isosurfaces of the wavelet coefficients in the wavelet analysis on a grid of $128\times 128$ pixels, with $a=64$ pixels corresponding to 1.0 Mpc. The images present decreasing wavelet scales:
$a=64,32,16,~{\rm and}, 8$, respectively.
\label{fig5}}
\end{figure} 

\begin{figure}
\includegraphics[width=8.5cm,height=8.5cm]{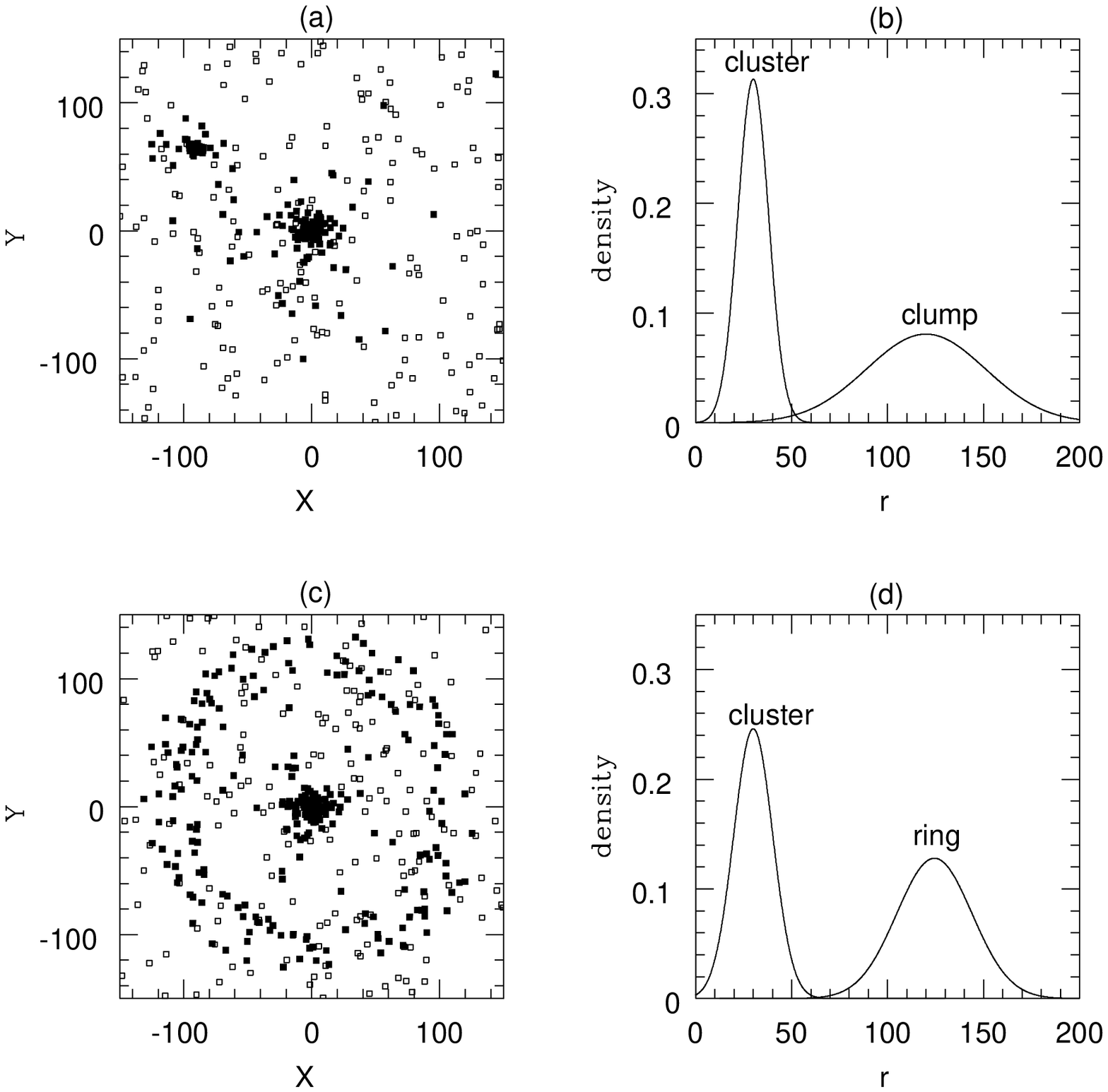}
\caption{Morphology test. Pannels (a) and (c) exhibit examples of
the distribution of points for the cases cluster+clump and cluster+ring, respectively. Pannels (b) and (d)
show the result of the spatial analysis: normal distributions related to the scales detected 
for both geometries.
\label{fig6}}
\end{figure}

\section{Anisotropy test}

Spheroidal clumps and rings (and possibly other
shapes) may produce approximately the same results after spatial analysis. However, in the case
of comparing rings and clumps, there is a remaining property to be tested: the anisotropy of the
signal around the cluster. This second test is now introduced. 

For the projected distribution, 
anisotropy can be probed by the reduced second-order moment measure ${\mathcal K}$
of a point pattern (e.g. Illian et al. 2008 and references therein). 
In this work, we estimate ${\mathcal K}$ with using the library {\bf spatstat}
under R statistical package. The routine {\bf Kmeasure} executes the following steps:

\begin{enumerate}

\item it takes a point pattern,
\item and forms the list of all pairs of distinct points in the pattern,
\item then computes the vectors that join
the first point to the second point in each pair,
\item and treats these vectors as a pattern of `points', 
\item finally applyies a Gaussian kernel
smoother to them (keeping the kernel width $30^{\prime\prime}$, see Sections 4 and 5).

\end{enumerate}

The algorithm approximates the point pattern and its window by
a binary pixel image, introduces a Gaussian smoothing kernel and
uses the Fast Fourier Transform to make a density estimate $\kappa$.
The density estimate of $\kappa$ is returned in the form of a
real-valued pixel image. The $\hat{\mathcal K}$ estimator is defined as the the expected number of points 
lying within a distance $r_{max}$ of a typical point, and with displacement 
vector having an orientation in the range $[\alpha, \beta]$. 
This can be computed by summing the entries over the relevant region, i.e, the
sector of the disc of radius $x$ centred at the origin with angular range
$[\alpha,\beta]$. Hence, we can compute a measure of anisotropy, A, as 
integrals of the form 

\begin{equation}
A\equiv\int_0^{r_{max}}\int_\alpha^\beta d\kappa(r,\theta).
\end{equation}

Recall that Ripley function (both in first and second order) is used to
test the hypothesis that a given planar point pattern is a realization
of a Poisson process. In the case of vectors instead of points, the
second-order orientation analysis can be done around the center of
the field. A nearly flat anisotropy profile would be consistent with a Poisson process $A\approx 0$.
Thus, the idea here is to look for significant anisotropies in integral (7) for angular steps 
${\rm\Delta\theta=\beta-\alpha}$. The choice of ${\rm \Delta\theta}$ is arbitrary, and it was set to be ${\rm \Delta\theta}=15^\circ$, based in the
following analysis. We take 1,000 realizations of a 
controlled sample corresponding to a point pattern
given by a Poisson distribution plus a Hernquist spheroid (Hernquist 1990)
located  at $\sim 100^{\prime\prime}$ and
making an angle of $45^\circ$ around the center. To justify our choice of
$\Delta\theta$ we present in Figure 7 the analyses for the case for
$\Delta\theta=5^\circ,15^\circ,~{\rm and}~30^\circ$.
For $\Delta\theta=5^\circ$, we still detect the peak, but now
there are secondary peaks and a more noisy behaviour for
$\langle A\rangle$. For  $\Delta\theta=30^\circ$,
the peak is still there, but now it is less significant.
This result suggests that in the limit of too small $\Delta\theta$ we have
a noisy anisotropy curve (possibly with false peaks), while
in the limit of very large $\Delta\theta$ the signal can be completely lost.
Actually, in the approximate range $[12^\circ,18^\circ]$ the resultant 
anisotropy profiles are almost indistinguishable, with relative differences  $\lesssim 3\%$.

\begin{figure}
\includegraphics[width=8cm,height=8cm]{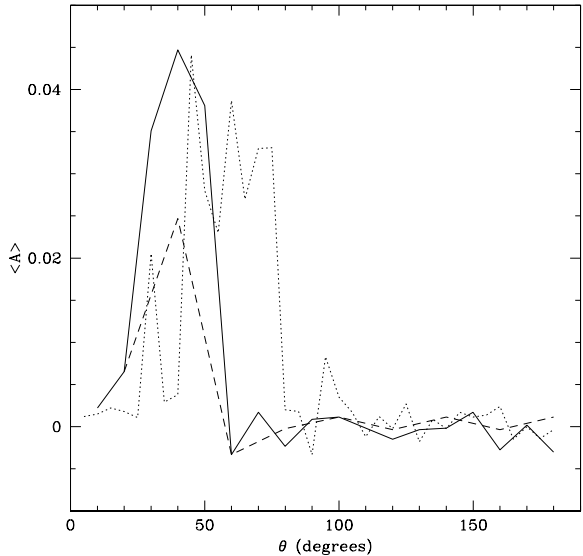}
\caption{Anisotropy profiles of a controlled sample (Poisson + clump at $45^\circ$).
Solid lines for $\Delta\theta=15^\circ$; dotted lines for $\Delta\theta=5^\circ$;
and dashed lines for $\Delta\theta=30^\circ$.
\label{fig}}
\end{figure}

Results of the anisotropy test for the averaged samples
are presented in Figure 8. Note that Model I (although very simplified) reproduces
fairly well the anisotropy profile of real data, exhibiting a peak around $110^\circ$ 
(or the supplementary angle $290^\circ$), a direction corresponding to the axis crossing
the NW quadrant. On the contrary, Model II has a flatter anisotropy profile.
In fact, Kolmogorov-Smirnov test is consistent with the null hypothesis  that Data and Model I samples 
come from the same distribution. This is not true for the KS test comparing Data and Model II samples,
and Model I and Model II samples (see p-values in Figure 8).
I also have applied a Welch two-sample t-test to find if the anisotropy profiles are stochastically higher
between each other. Again, the results indicate that both Data and Model I have profiles significantly 
higher than that of Model II. Also, there is not a significant difference between 
Data and Model I profiles (see p-values in Figure 8). This all means that the galaxy field around
Cl 0024+17 is more consistent with an anisotropic cluster+clump distribution instead of a cluster+ring one.

Morphology and anisotropy tests are not totally conclusive, but they suggest that the second scale found
in data is more probable to be associated to the clump at $\sim 120^{\prime\prime}$
NW from the cluster center than to a ring
of galaxies, although this feature cannot be discarded by spatial analysis alone.

\begin{figure}
\includegraphics[width=8cm,height=8cm]{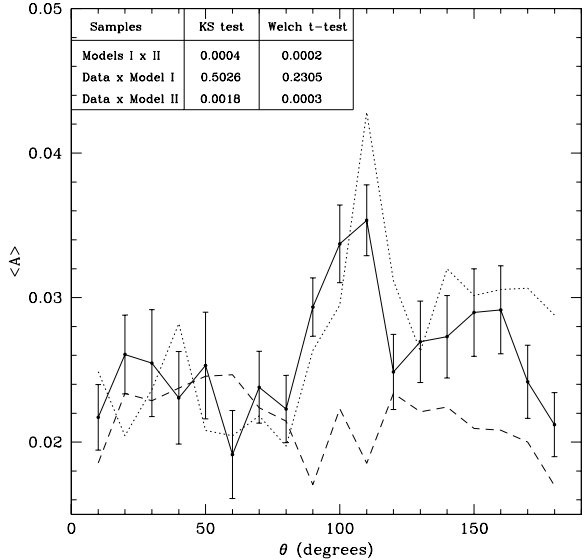}
\caption{Anisotropy test. The anisotropy profiles for the averaged samples are presented as follows:
Data -- solid lines; Model I -- dotted lines; Model II -- dashed lines.
p-values associated to KS test and Welch t-test are presented in the
upper left corner table.
\label{fig7}}
\end{figure}

\section{Summary and Discussion}

By considering the planar galaxy distribution around Cl 0024+17 as a SPP,
I have applied some tools of spatial statistics analysis to probe the existence
of an extended structured in this cluster.  The catalog was corrected
through 1,000 MC  realizations of original the sample, following the
completeness map of \cite{czo1}.
Based on the Besag L-function curves
I have identified two distinct physical scales where the null hypothesis that the data came of
a Poisson process is rejected at 95\% confidence. Values found for the two scales present 
normal distributions: $\mu_1=31.4^{\prime\prime}; \sigma_1=8.4^{\prime\prime}$ and  
$\mu_1=112.9^{\prime\prime}; \sigma_2=21.4^{\prime\prime}$.
Both Bartlett and Kruskal-Wallis statistical tests indicate these normals are 
significantly unequal both in variances and means. This reinforces the idea of two
density peaks in the Cl 0024+17 field. Also, the analysis of the averaged
correlation function for all MC realizations suggests that at $\sim 50^{\prime\prime}$
may be the maximum scale for the core of the system. This scale is approximately
at the intersection of the normal distributions found for the two transition
scales. The DM ring radius and the
$2^{\rm nd}$ transition scale are approximately coincidents with the highest peaks
of $\xi (r)$ after $\sim 50^{\prime\prime}$. However, morphology and anisotropy tests
based on simulations of 1,000 mock fields for cluster+clump and cluster+ring models
are not conclusive about interpreting the second scale as a ring of galaxies.
However, a wavelet analysis of the field points out that a cluster+clump model
cannot be rejected at scales $a=32$ pixels (0.50 Mpc) and $a=16$ pixels (0.25 Mpc).
Finally, the anisotropy test also suggests that the second scale found
in data is more probable to be associated to the clump at $\sim 120^{\prime\prime}$
NW from the cluster center than to a ring
of galaxies.

Summing up this work, the analyses presented in Sections 4, 5 and 6 are not
indicative that galaxies follow dark matter around Cl 0024+17, but the methodology
based on spatial patterns examination seems to have good sensitivity to find
physical scales in galaxy systems (both in radial and angular
coordinates). It could turn a useful tool to study galaxy (and maybe other
astrophysical) systems in general. Spatial analysis (followed by subsidiary morphology and anisotropy tests)
could be a poweful technique to probe remnant substructures of clusters collisions using
galaxy data alone. Analysing a large dataset of galaxy clusters, we could identify
candidates to bullet-type clusters by the presence of at least two transition scales.
Also, the approach can introduce a new estimator for galaxy
systems radii based only on the scales detected in data \cite{riblop}.
Finally, this method can be directly applied to data from N-body simulations of
cluster collisions. Some sensitivity to initial conditions could be tested
not only looking for bumps in  the density projected along the
collision axis, but also with the dark matter particles interaction range.
This is an interesting test to be done, since a recent study
suggests that the ring-like feature of Cl 0024+17 (and similar cases) 
would be associated to very improbable initial velocity distributions \cite{zu}.

\section{Acknowledgments}
I thank the referee for very useful suggestions.
I am grateful to A.C. Schilling and B. Carvalho for helpful statistical discussions.
I also thank the support of CNPq, under grant 201322/2007-2.
Finally, I thank the Fermilab for the hospitality.

\end{document}